# Tuning the morphology and magnetic properties of single-domain $SrFe_8Al_4O_{19}$ particles prepared by citrate auto-combustion route

Anastasia E. Sleptsova[a], Liudmila N. Alyabyeva[b], Evgeny A. Gorbachev*[a,c], Ekaterina S. Kozlyakova[d], Maxim A. Karpov[a], Chen Xinming[c], Alexander V. Vasiliev[a], Boris P. Gorshunov[b], Anatoly S. Prokhorov[b,e], Pavel E. Kazin[a], Lev A. Trusov[a,c]

[a] Department of Chemistry, Lomonosov Moscow State University, Moscow, 119991, Russia.

e-mail: ev.a.gorbachev@gmail.com.

[b] Laboratory of Terahertz Spectroscopy, Center for Photonics and 2D Materials, Moscow Institute of Physics and Technology, Dolgoprudny, 141701, Russia

[c] MSU-BIT University, Shenzhen, 517182, China

[d] Department of Physics, Lomonosov Moscow State University, Moscow, 119991, Russia

[e] Prokhorov General Physics Institute of the Russian Academy of Sciences, Moscow 119991, Russia

ABSTRACT: Single-domain particles of $SrFe_8Al_4O_{19}$ were prepared by thermal treatment at 1200°C of porous products of citrate-nitrate auto-combustion, and the influence of synthesis time on the particle morphology and magnetic properties was studied. The procedure allows to obtain $SrFe_8Al_4O_{19}$ particles with mean diameters 100 – 460 nm, and their coercivity ranges from 14.5 to 18.4 kOe, while ferromagnetic resonance frequencies vary from 149 to 164 GHz.

Although M-type hexaferrites ($BaFe_{12}O_{19}$ and $SrFe_{12}O_{19}$) are well-studied compounds, they are still of great interest due to unique set of finely tunable functional properties[1]. Large magnetocrystalline anisotropy and high coercivity form the basis for their wide applications as permanent magnets[2], and high thermal and chemical stability make it possible to produce nanomagnets for magnetic recording media[3–5], fast-response magneto-active colloids[6,7] and hard magnetic cores of exchange-coupled nanocomposites[8,9]. Moreover, the hexaferrites display specific millimeter-wave (sub-THz) absorption due to ferromagnetic resonance (FMR), which is essential for modern wireless communication technologies[10–13]. The properties of the hexaferrites are extremely sensitive to particles morphology as well as to ionic substitutions in their crystal structure.

Recently, we presented a method for manufacturing highly aluminum substituted strontium hexaferrite in the form of single-domain particles[13,14]. This resulted in hexaferrite materials with giant coercivity up to 40 kOe and record-high natural FMR (under zero magnetic field) frequencies 160 – 250 GHz. The method itself is quite simple, and it is based on heat treatment of porous oxide precursors obtained by citrate auto-combustion route. Nevertheless, it implies the possibility to modify the particle morphology and functional properties by varying heat treatment parameters.

Herein, we study the influence of the annealing time at a temperature of 1200°C on the $SrFe_8Al_4O_{19}$ particles morphology, magnetic properties, and millimeter-wave absorption.

M-type hexaferrite particles with composition $SrFe_8Al_4O_{19}$ were produced by citrate route reported previously in [14]. Briefly, strontium, iron and aluminum nitrates and citric acid (all high purity grade from Sigma-Aldrich) were mixed in an aqueous solution to obtain a molar ratio of metal ions to citrate ion of 1:3. The solution was neutralized by $NH_3(aq)$ and then dehydrated by heating in a sand bath. The product was spontaneously combusted to form a highly porous precursor powder. The powder was heated to 1200°C with a rate of 10 °C min$^{-1}$ and exposed at this temperature for 0, 0.5, 2, 8, 14 and 24 hours.

According to X-ray diffraction (Rigaku D-Max 2500, Cu-K$_\alpha$ radiation), all the obtained samples represent single crystalline M-type hexaferrite phase. The lattice parameters (Table 1) of the sample exposed for 0 h are slightly larger, than these of the hexaferrite exposed for longer times, indicating that the Al substitution is not fully completed. The hexaferrite composition estimated from the lattice parameters is $SrFe_{8.15}Al_{3.85}O_{19}$[14]. The parameters of samples exposed for 0.5 h and longer are in a good agreement with previously reported for $SrFe_8Al_4O_{19}$ phase.[14]

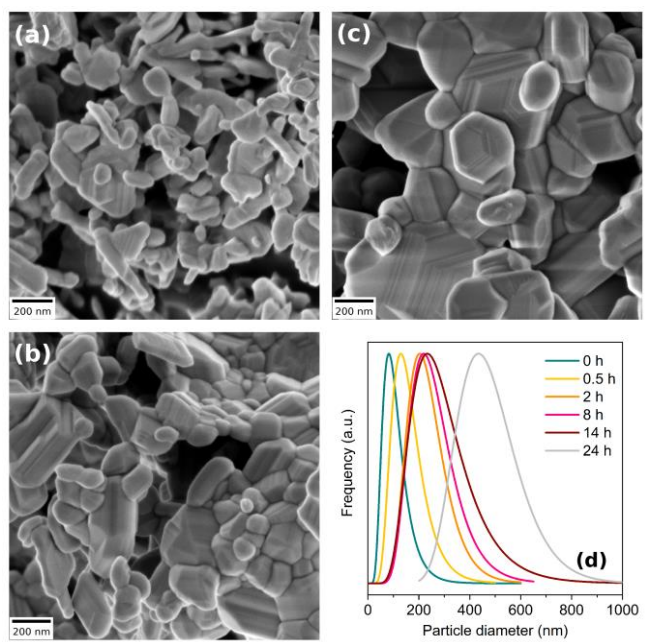

**Figure 1.** a) – c) SEM images of hexaferrite powders obtained at 1200 °C with exposiure times of 0 h, 0.5 h and 14 h, respectively; d) particle diameter lognormal distributions calculated by fitting histogramms.

The particle morphology (Fig.1 a-c, Table 1) was determined by scanning electron microscopy (Carl Zeiss NVision40). The particles have mostly a thick-plate shape and reveal wide diameter distributions (Fig.1 d). The mean particle diameter rises with an increase in exposure time from 100 to 460 nm for 0 h and 24 h, respectively. The estimated single-domain limit of $SrFe_8Al_4O_{19}$ is supposed to be 8 μm[13], therefore the most of the particles within the samples must be in the single-domain state.

**Table 1.** Unit cell parameters ($a$, $c$), $x$ – aluminum content estimated from the lattice parameters, $d_{mean}$ – mean particle diameter, magnetic properties ($H_C$ – coercive filed, $M_S$ – saturation magnetization at 60 kOe, $M_r$ – remanence), $f_r$ – natural ferromagnetic resonance frequency.

| Exposure time, h | $x$ | $a$, Å | $c$, Å | $d_{mean}$, nm | $H_C$, kOe | $M_S$, emu/g | $M_r$, emu/g | $M_r/M_S$ | $f_r$, GHz |
|---|---|---|---|---|---|---|---|---|---|
| 0 | 3.85 | 5.7905(1) | 22.7459(7) | 100 | 14.5 | 14.0 | 7.5 | 0.51 | 149 |
| 0.5 | 3.95 | 5.7886(1) | 22.7330(7) | 150 | 15.3 | 13.4 | 7.3 | 0.52 | 155 |
| 2 | 4 | 5.7880(1) | 22.7311(4) | 230 | 15.6 | 13.5 | 7.3 | 0.52 | 164 |
| 8 | 4 | 5.7874(1) | 22.7283(5) | 260 | 15.9 | 13.5 | 7.4 | 0.53 | 164 |
| 14 | 4 | 5.7879(1) | 22.7277(5) | 280 | 16.8 | 13.4 | 7.1 | 0.51 | 164 |
| 24 | 4 | 5.7878(1) | 22.7280(4) | 460 | 18.4 | 13.6 | 7.3 | 0.51 | 164 |

The magnetic hysteresis loops of the samples (Quantum Design PPMS, magnetic fields up to 6 T) are shown in Fig.2 and the corresponding properties are summarized in Table 1. The shapes of the hysteresis loops are typical for randomly oriented single-domain Stoner-Wohlfarth particles with $M_r/M_S$ close to 0.5.[15] The substitution of paramagnetic iron ions by diamagnetic aluminum in hexaferrite structure results in gradual decrease of saturation magnetization $M_S$[1,16], while the coercivity rises only for single-domain particles[14]. The saturation magnetization ($M_S$) of the samples annealed for 0.5 – 24 h is close to the reported previously for $SrFe_8Al_4O_{19}$[14]. The 0 h sample has higher $M_S$, which indicates lower Al substitution, and it is in an accord with XRD analysis. The coercivity sharply jumps up with an increase in time of exposure from 0 h to 0.5 h, which also is the result of higher aluminum substitution. For annealing times 0.5 – 24 h, the coercivity gradually rises from 15.3 to 18.4 kOe. While the coercivity is strongly affected by particle size and crystallinity[1,17], the increase of the coercivity is due to continuous recrystallization resulting in particle growth.

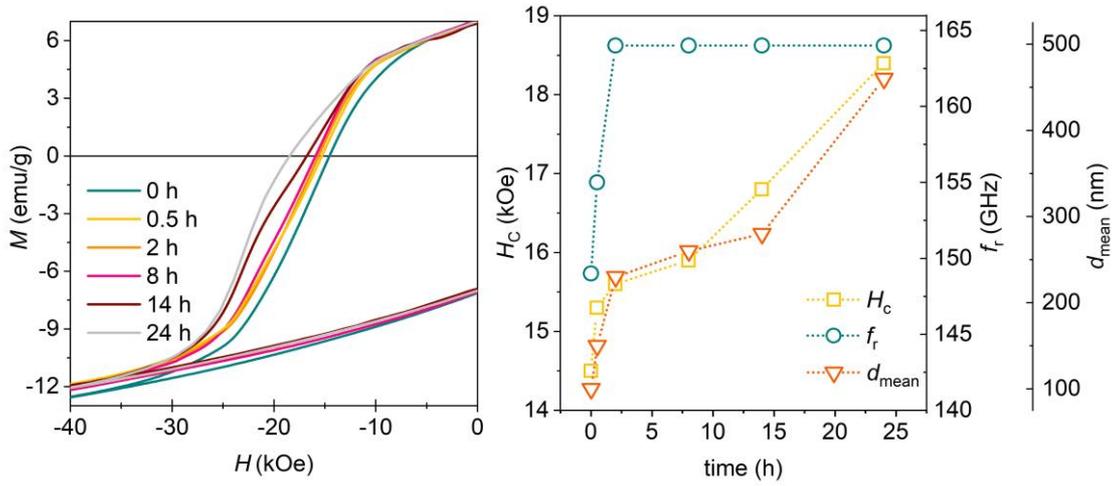

**Figure 2.** Left: Magnetic hysteresis loops. Right: dependence of coercivity ($H_C$), ferromagnetic resonance frequency ($f_r$) and mean particle diameter ($d_{mean}$) on exposure time at 1200 °C.

Room temperature FMR spectra of the samples (composite pellets of PMMA/hexaferrite powder (1:9 wt ratio) fabricated via a hot-pressing as described in our previous work [13]) were recorded with a terahertz time-domain spectrometer Teraview TPS 3000 in absence of external magnetic field. The samples possess FMR frequencies at 149 GHz for 0 h, 155 GHz for 0.5 h and 164 GHz for the rest of exposure times resulting the similar value for $MFe_8Al_4O_{19}$ compound[13]. As one can see, $f_r$ value is sensitive only to hexaferrite composition while particle size does not affect it. This confirms the Kittel's formula $f_r \sim H_a$[18], where anisotropy field $H_a$ is known to be strongly depended on an aluminum concentration while it is indifferent to a particles size.

In conclusion, we studied the formation of single-phase $SrFe_8Al_4O_{19}$ hexaferrite via annealing porous products obtained by auto-combustion of citrate solutions. The described method offers a promising way to prepare particles of highly substituted hexaferrites with tunable particles size as well as high magnetic and millimeter-wave absorption properties. The value of 14.5 kOe is the highest reported coercivity of hexaferrite nanosized particles. It was also found that mean particle diameter in the range 230 – 460 nm does not affect the ferromagnetic resonance frequency of the $SrFe_8Al_4O_{19}$ hexaferrites that is 164 GHz, however it has a strong effect on their coercivity, which increases from 15.3 up to 18.4 kOe, respectively. The reported hexaferrite powders with high phase purity and tunable particle size within the single-domain region are essential for production of fine-grained ceramics, hard magnetic films, coatings, and composites. Due to their high-frequency absorption properties and high magnetic hardness the developed hexaferrite materials are prospective for modern applications, such as spintronics, electromagnetic shielding, durable magnetic recording, and the next generation of wireless technologies.

This work was supported by the Russian Foundation for Basic Research (grant no. 20-02-00887). SEM measurements were performed using the equipment of the JRC PMR IGIC RAS.